\documentclass[twocolumn,english,aip,jcp,numerical,reprint,10pt]{revtex4-1}
\usepackage[T1]{fontenc}
\usepackage[latin9]{inputenc}
\usepackage{amsmath}
\usepackage{amssymb}
\usepackage{graphicx}
\usepackage{esint}

\makeatletter

\newcommand{\lyxdot}{.}

\PassOptionsToPackage{caption=false}{subfig}
\usepackage{braket}

\usepackage{babel}

\makeatother

\usepackage{babel}
\begin{document}

\title{On the accuracy of surface hopping dynamics in condensed phase non-adiabatic
problems}

\author{Hsing-Ta Chen}

\affiliation{Department of Chemistry, Columbia University, New York, New York
10027, U.S.A.}

\author{David R. Reichman}

\affiliation{Department of Chemistry, Columbia University, New York, New York
10027, U.S.A.}
\begin{abstract}
We perform extensive benchmark comparisons of surface hopping dynamics
with numerically exact calculations for the spin-boson model over
a wide range of energetic and coupling parameters as well as temperature.
We find that deviations from golden-rule scaling in the Marcus regime
are generally small and depend sensitively on the energetic bias between
electronic states. Fewest switched surface hopping (FSSH) is found
to be surprisingly accurate over a large swath of parameter space.
The inclusion of decoherence corrections via the augmented FSSH (A-FSSH)
algorithm improves the accuracy of dynamical behavior compared to
exact simulations, but the effects are generally not dramatic, at
least for the case of an environment modeled with the commonly used
Debye spectral density. 
\end{abstract}
\maketitle

\section{Introduction}

Electronically non-adiabatic transitions lie at the heart of some
of the most important dynamical processes in the physical sciences.\cite{Tully2012}
Phenomena ranging from gas phase atomic and molecular collisions\cite{Miller}
to electron and energy transfer in the condensed phase\cite{Nitzan}
are often intimately influenced by the coupling between distinct potential
energy surfaces that is induced by nuclear motion. Theoretically,
the accurate treatment of non-adiabatic dynamics is challenging, in
particular in condensed phase applications where the interplay between
the large number of nuclear degrees of freedom with multiple coupled
electronic states greatly increases the complexity of the problem.
Exact results may be obtained for specific idealized models such as
spin-boson systems where potential energy surfaces are harmonic and
linearly displaced.\cite{Tanimura1989,Makarov1994a,Makri1995,Thoss2001}
In more realistic condensed phase situations, exact solutions are
currently out of reach, despite much recent progress.\cite{Ben-Nun2000}

Among the myriad approximate methods for treating non-adiabatic dynamics,
the surface hopping approach\cite{Tully1971,Tully1990,Tully1998,Barbatti2011}
stands out for several reasons. First, the method is equally applicable
to gas phase and condensed phase problems, and can be used to treat
realistic anharmonic nuclear motion on potential energy surfaces,
albeit in a classical manner.\cite{Hammes-Schiffer1994,Hammes-Schiffer1995,Kim2006}
Surface hopping has the advantage that it is naturally formulated
in the adiabatic picture, so that it can conveniently be employed
in conjunction with electronic structure calculations. The method
is also inexpensive, non-perturbative, and provides a superior description
of branching processes and detailed balance when compared to other
approaches, such as the Ehrenfest method.\cite{Hsieh2013a} Despite
these appealing features, surface hopping naturally suffers from several
deficiencies.\cite{Muller1997,Topaler1998,Hack2001,Subotnik2011c}
Clearly the description of nuclear motion as classical renders the
approach incapable of capturing low temperature effects such as nuclear
tunneling on a single potential surface.\cite{Ben-Nun2007,Billeter2001,Agarwal2002}
More generally, while surface hopping does not employ perturbation
theory in any parameter, as with nearly all mixed quantum-classical
approaches to non-adiabatic dynamics, it cannot be \textit{systematically}
derived from an exact starting point.\cite{Kapral1999,Subotnik2013}$^{,}$\footnote{Clearly, Ref.~\onlinecite{Kapral1999} and \onlinecite{Subotnik2013}
outline steps towards a complete derivation of the FSSH algorithm
starting from the exact equation of motion for the density matrix.
In each case, however, there is at least one step that needs to be
assumed for which the domain of validity is difficult to assess. It
is in this strict sense that we refer to the lack of a systematic
derivation of FSSH.} This fact makes it difficult to evaluate surface hopping's domain
of validity.

One long recognized shortcoming of surface hopping is the fact that,
in its standard implementation, the algorithm does not provide decoherence
for electronic amplitudes. This knowledge has led to the development
of important modifications of surface hopping aimed at more accurately
describing decoherence.\cite{Webster1994,Schwartz1996,Wong2002,Wong2002a,Bedard-Hearn2005,Larsen2006,Zhu2004,Zhu2004a,Jasper2005,Horenko2002,Shenvi2011a,Fang1999,Fang1999a,Prezhdo1998a}
In an important recent series of studies, Landry and Subotnik showed
that a striking consequence of the neglect of decoherence in surface
hopping is the failure to properly capture the golden-rule scaling
of the non-adiabatic transfer rate in the Marcus regime.\cite{Landry2011,Subotnik2011}
It should be noted, however, that there has been some debate as to
just how pervasive this problem is.\cite{Jiang2012b,Wang2013a,Schwerdtfeger2014,Falk2014,Jain2015b}
One of the goals of the present work is to provide an in depth examination
of this issue.

More broadly we aim to compare surface hopping, with and without corrections
for decoherence, to exact calculations in a model condensed phase
system, namely the spin-boson model.\cite{Leggett1987,MacKernan2002}
Although the spin-boson model is an idealized proxy for a real condensed
phase system exhibiting non-adiabatic transitions, it offers the advantage
that algorithms now exist that enable the calculation of exact dynamics
over a wide swath of the relevant parameter space.\cite{Tanimura1989,Makarov1994a,Makri1995,Thoss2001}
While in the past surface hopping was compared to exact benchmark
calculations of low dimensional scattering problems,\cite{Tully1990,Tully1998,Shenvi2011a}
we now can provide guidelines for understanding the successes and
failures of the surface hopping approach in a broader condensed phase
setting. It should be noted, however, that we will restrict our comparison
to the \textquotedbl{}overdamped\textquotedbl{} case of the coupling
to a Debye spectral density, since it is here that facile exact simulations
may be performed. While the Debye case represents perhaps the most
commonly employed model of a condensed environment in the spin-boson
context, our choice implies that some aspects related to the interplay
between surface hopping trajectories and decoherence which are expected
to be most dramatic and subtle in the underdamped limit, may not arise.\cite{Zhu2004a,Jasper2004,Jain2015b}
Regardless, our work should at least provide a starting point for
assessing how surface hopping performs in generic condensed phase
settings.

Our paper is organized as follows: We begin in Sec.~\ref{sec:Fewest-Switch-Surface}
with a review of the standard surface hopping algorithm for the spin-boson
model and various formulations of decoherence corrections. In Sec.~
\ref{sec:The-Golden-Rule}, we present our results for the scaling
of the non-adiabatic transfer rate with respect to the electronic
coupling in the golden-rule regime. In Sec.~\ref{sec:the-full-parameter},
we explore the full parameter space of spin-boson model. We summarize
our results and conclude in Sec.~\ref{sec:Conclusions}.

\section{Fewest Switch Surface Hopping (FSSH) and Decoherence\label{sec:Fewest-Switch-Surface}}

\subsection{Spin-Boson Model}

We consider the spin-boson model, $H=H_{s}+H_{b}+H_{sb}$, which describes
a two-level system with energy bias $\epsilon_{0}$ and constant diabatic
coupling $\Delta$ 
\begin{equation}
H_{s}=\epsilon_{0}\sigma_{z}+\Delta\sigma_{x},
\end{equation}
interacting with an infinite set of harmonic oscillators (bath) 
\begin{equation}
H_{b}=\sum_{j}\frac{1}{2}\left(p_{j}^{2}+\omega_{j}^{2}q_{j}^{2}\right),
\end{equation}
where $\omega_{j}$ is the frequency of the $j$-th bath mode.\cite{weiss1999}
The isolated electronic system and the bath are coupled bilinearly
\begin{equation}
H_{sb}=\sigma_{z}\sum_{j}g_{j}q_{j},
\end{equation}
where $g_{j}$ is the coupling strength between the two-level system
and the $j$-th harmonic oscillator. We adopt the Pauli matrix notation
$\sigma_{x}=\left|1\right\rangle \left\langle 2\right|+\left|2\right\rangle \left\langle 1\right|$
and $\sigma_{z}=\left|1\right\rangle \left\langle 1\right|-\left|2\right\rangle \left\langle 2\right|$
where $\left|i\right\rangle $ indicate the diabatic states of the
system. Throughout the present paper, we use mass scaled coordinates
and momenta for the bath modes, $q_{j}=\sqrt{M_{j}}Q_{j}$ and $p_{j}=P_{j}/\sqrt{M_{j}}$,
where $M_{j}$ are the effective mass of nucleus for the $j$-th harmonic
oscillator and set $\hbar=1$. We denote bold letters $\boldsymbol{q}$,
$\boldsymbol{p}$ by the vector of nuclear degrees of freedom.

The influence of the bath on the dynamics of the system can be captured
in the compact form of a spectral density, 
\begin{equation}
J\left(\omega\right)=\frac{\pi}{2}\sum_{j}\frac{g_{j}^{2}}{\omega_{j}}\delta\left(\omega-\omega_{j}\right).
\end{equation}
In the present paper, we consider the Debye model of the spectral
density,\cite{Thoss2001} 
\begin{equation}
J\left(\omega\right)=\frac{E_{r}}{2}\frac{\omega\omega_{c}}{\omega^{2}+\omega_{c}^{2}},
\end{equation}
which is appropriate for the description of a solvent environment
with Debye dielectric relaxation. The Debye spectral density function
is characterized by two parameters, the reorganization energy $E_{r}$,
and the characteristic bath frequency $\omega_{c}$. In electron-transfer
theory, the reorganization energy represents a direct measure of the
coupling strength between the system and the bath. The characteristic
frequency is related to the relaxation time scale of the bath, $\tau=1/\omega_{c}$.
The Debye spectral density spans broader frequency than the standard
Ohmic ($J(\omega)\propto\omega e^{-\omega/\omega_{c}}$) and Brownian
forms ($J(\omega)\propto\omega/((\omega^{2}-\omega_{c}^{2})^{2}+\gamma^{2}\omega^{2})$).
Following the procedure outlined in Refs.~\onlinecite{Rekik2013,
Wang2001}, it is convenient to discretize the Debye spectral density
function via $\omega_{j}=\tan\left(\left(j-0.5\right)\tan^{-1}\left(\omega_{\text{max}}/\omega_{c}\right)/N\right)$
where $\omega_{\text{max}}$ is the largest frequency and $N$ is
the number of oscillators employed in the discretization.

The population dynamics of the spin-boson model can be calculated
by the numerically exact hierarchical equations of motion (HEOM) methodology,\cite{Tanimura1989}
implemented in the Parallel Hierarchy Integrator (\texttt{PHI}).\cite{Strumpfer2012}
The HEOM method is easier to use when the spectral density take the
Debye form so that the bath correlation function can be written as
a sum of exponentially decaying functions in time.\cite{may2008charge,Shi2003}
We use the HEOM method to produce all of our benchmark results for
the spin-boson model.

We focus on the reduced population dynamics of the system

\begin{equation}
P_{i}\left(t\right)=\text{Tr}_{b}\left\{ \rho\left(0\right)e^{iHt}\left|i\right\rangle \left\langle i\right|e^{-iHt}\right\} 
\end{equation}
where we assume a factorized initial condition $\rho\left(0\right)=\rho_{b}\left|1\right\rangle \left\langle 1\right|$
and 
\begin{equation}
\rho_{b}=\frac{e^{-\beta H_{b}}}{\text{Tr}_{b}\left\{ e^{-\beta H_{b}}\right\} },
\end{equation}
with the inverse temperature of the bath, $\beta=1/kT$. The initial
condition of the system corresponds to an impulsive Franck-Condon
transition with the bath in a state independent of the system with
oscillators centered at $q_{j}=0$.

\subsection{FSSH and its variants}

The fewest-switches surface hopping (FSSH) algorithm is a mixed quantum-classical
method that treats the bath degrees of freedom classically and the
electronic system quantum mechanically.\cite{Tully1971,Tully1990,Tully1998}
A swarm of classical nuclear trajectories evolve on the adiabatic
potential energy surfaces associated with the electronic states with
each \textit{individual} trajectory evolving on a single \textit{active}
surface. Along each trajectory, the electronic wave function propagates
according to the Schrodinger equation with the classical nuclear variables
evolving as parameters. The essence of FSSH is to simulate the population
of the electronic states via the density of trajectories on each surface.
For this purpose, a surface-hopping scheme is introduced to allow
trajectories to hop among the adiabatic energy surfaces and match
the electronic populations. The hopping probability of the classical
bath trajectories depends on the electronic wave functions with specific
conditions for the acceptance of non-adiabatic transitions. Instead
of listing these conditions, we describe them within the context of
the spin-boson model.

To implement the FSSH algorithm for the spin-boson model, we transform
the model to its adiabatic representation by diagonalizing the Hamiltonian
$H|\Phi_{i}\left(\boldsymbol{q}\right)\rangle=(\frac{\boldsymbol{p}^{2}}{2}+V_{i}\left(\boldsymbol{q}\right))|\Phi_{i}\left(\boldsymbol{q}\right)\rangle$
where 
\begin{equation}
V_{i}\left(\boldsymbol{q}\right)=\frac{1}{2}\sum_{j}\omega_{j}^{2}q_{j}^{2}+\left(-1\right)^{i}\sqrt{\left(\boldsymbol{g\cdot q}+\epsilon_{0}\right)^{2}+\Delta^{2}}
\end{equation}
are the adiabatic potential energy surfaces and $\boldsymbol{g\cdot q}=\sum_{j}g_{j}q_{j}$.
One may transform the diabatic states to the adiabatic representation
via the unitary transformation $|\Phi_{i}\left(\boldsymbol{q}\right)\rangle=\sum_{j}U_{ij}\left(\boldsymbol{q}\right)|j\rangle$
where 
\begin{equation}
U\left(\boldsymbol{q}\right)=\left(\begin{array}{cc}
\sin\theta\left(\boldsymbol{q}\right) & -\cos\theta\left(\boldsymbol{q}\right)\\
\cos\theta\left(\boldsymbol{q}\right) & \sin\theta\left(\boldsymbol{q}\right)
\end{array}\right).
\end{equation}
The adiabatic-diabatic mixing angle is defined as $\theta\left(\boldsymbol{q}\right)=\frac{1}{2}\tan^{-1}\left(\Delta/(\boldsymbol{g\cdot q}+\epsilon_{0})\right)$
which depends on the bath coordinates. Within the adiabatic representation,
the electronic wavefunction can be written as $|\Psi\left(t\right)\rangle=c_{1}\left(t\right)|\Phi_{1}\left(\boldsymbol{q}\right)\rangle+c_{2}\left(t\right)|\Phi_{2}\left(\boldsymbol{q}\right)\rangle$
and the adiabatic amplitudes satisfy an implicit time-dependent Schrodinger
equation 
\begin{equation}
\frac{d}{dt}c_{i}\left(t\right)=-iV_{i}\left(\boldsymbol{q}\right)c_{i}\left(t\right)-\sum_{k}\boldsymbol{p\cdot d}_{ik}\left(\boldsymbol{q}\right)c_{k}\left(t\right),
\end{equation}
where $d_{ik}^{j}\equiv\left\langle \Phi_{i}\left(\boldsymbol{q}\right)\right|\frac{d}{dq_{j}}\left|\Phi_{k}\left(\boldsymbol{q}\right)\right\rangle $
is the derivative coupling matrix. For the spin-boson model, the derivative
coupling matrix elements are $d_{11}^{j}=d_{22}^{j}=0$ and 
\begin{equation}
d_{12}^{j}=-d_{21}^{j}=\frac{g_{j}}{2}\frac{\Delta}{\left(\boldsymbol{g\cdot q}+\epsilon_{0}\right){}^{2}+\Delta^{2}}.
\end{equation}
We define the pure state electronic density matrix $\hat{\sigma}$
by $\sigma_{ik}=c_{i}c_{k}^{*}$ and the equivalent equation for the
density matrix can be written as 
\begin{equation}
\frac{d}{dt}\hat{\sigma}\left(t\right)=-i\left[\hat{V}\left(\boldsymbol{q}\right),\hat{\sigma}\left(t\right)\right]-\left[\boldsymbol{p\cdot}\hat{\boldsymbol{d}}\left(\boldsymbol{q}\right),\hat{\sigma}\left(t\right)\right],
\end{equation}
where the potential energy matrix is $V_{ik}\left(\boldsymbol{q}\right)=\delta_{ik}V_{i}\left(\boldsymbol{q}\right)$.

The bath in the FSSH algorithm is described via a swarm of trajectories
evolving classically on adiabatic potential surfaces. Each individual
trajectory propagates on\textit{ }the\textit{ active} adiabatic potential
surface, $V_{a}\left(\boldsymbol{q}\right)$, via $\boldsymbol{\dot{q}=p}$
and $\dot{\boldsymbol{p}}=-\partial V_{a}/\partial\boldsymbol{q}$,
and the bath configuration is followed by monitoring the time-dependence
of $\left(\boldsymbol{q}^{(n)},\boldsymbol{p}^{(n)},a^{(n)}\right)$
for $n=1,\cdots,N_{\text{traj}}$. Each trajectory is allowed to switch
active surfaces in order to force the relative number of trajectories
on each surface to mimic the adiabatic probability calculated by the
adiabatic amplitudes. To accomplish this, a minimal switching probability
for a hop from surface $a$ (active) to surface $b$ (other) during
each time step $dt$ may be employed as\cite{Tully1990} 
\begin{equation}
\gamma_{ab}^{\text{hop}}=dt\frac{2}{|c_{a}|^{2}}\left[\text{Im}\left(V_{ba}\left(\boldsymbol{q}\right)c_{a}c_{b}^{*}\right)+\text{Re}(\boldsymbol{p\cdot d}_{ab}c_{a}c_{b}^{*})\right].\label{eq:hopping_rate}
\end{equation}
For the spin-boson model, the hopping probability is determined entirely
by the derivative coupling and the adiabatic coherence $c_{a}c_{b}^{*}$.
In addition to the hopping probability, trajectories must have enough
energy to hop to a new surface and obey energy conservation. If the
trajectory switches to a new active surface, the momentum is rescaled
in the direction of the derivative coupling by $\boldsymbol{p}'=\boldsymbol{p}+\kappa\boldsymbol{d}_{ab}$
satisfying $|\boldsymbol{p}+\kappa\boldsymbol{d}_{ab}|^{2}+2V_{b}\left(\boldsymbol{q}\right)=|\boldsymbol{p}|^{2}+2V_{a}\left(\boldsymbol{q}\right)$.

At time $t=0$, we require that the initial configuration of the bath
mimics the initial electronic density in the adiabatic representation.
The initial configuration for the bath modes are sampled from the
thermal Wigner distribution, $\rho_{b}\propto\exp\{-\sum_{j}\frac{2}{\omega_{j}}\tanh(\frac{\beta\omega_{j}}{2})(\frac{1}{2}p_{0j}^{2}+\frac{1}{2}\omega_{j}^{2}q_{0j}^{2})\}$,
with the trace over the bath approximated as $\text{Tr}_{b}\left\{ \rho_{b}\cdots\right\} \approx\frac{1}{N_{\text{traj}}}\sum_{(\boldsymbol{q}_{0},\boldsymbol{p}_{0})}^{w}\cdots\equiv\left\langle \cdots\right\rangle $.
In addition, we initialize the active configuration $a^{(n)}$ accordingly
by distributing the initial phase terms on surface $1$ with the probability
$|c_{1}\left(0\right)|^{2}$ and on surface $2$ via probability $|c_{2}\left(0\right)|^{2}$.

Given that the electronic amplitudes are propagated in the adiabatic
representation and the bath trajectories move along adiabatic energy
surfaces according to the FSSH algorithm, it is non-trivial to extract
diabatic electronic populations. We adopt the interpretation of mixed
quantum-classical density matrix\cite{Landry2013} for the diabatic
population on state $i$, which is given by 
\begin{align}
P_{i}= & \left\langle \sum_{j}\left|U_{ij}\left(\boldsymbol{q}\right)\right|^{2}\delta_{ja}\right.\nonumber \\
 & \left.+\sum_{j<k}2\text{Re}\left[U_{ij}\left(\boldsymbol{q}\right)\sigma_{jk}U_{ik}^{*}\left(\boldsymbol{q}\right)\right]\right\rangle .
\end{align}
Note that the expression for $P_{i}$ includes information from the
active surface ($a$) as well as the adiabatic amplitude ($\sigma_{jk}$).
For the spin-boson model, we can express the reduced population dynamics
of state $1$ as 
\begin{align}
P_{1} & =\left\langle \sin^{2}\theta\left(\boldsymbol{q}\right)\delta_{1a}+\cos^{2}\theta\left(\boldsymbol{q}\right)\delta_{2a}\right\rangle \nonumber \\
 & \qquad+\left\langle 2\sin\theta\left(\boldsymbol{q}\right)\cos\theta\left(\boldsymbol{q}\right)\text{Re}\left[c_{1}c_{2}^{*}\right]\right\rangle ,\label{eq:P1_expression}
\end{align}
which is composed of a portion associated with the active surface
and a portion contributed by the adiabatic coherence.

\subsection{Decoherence}

Within the standard FSSH algorithm, a difficulty arises when a trajectory
passes through the coupling region and the electronic wavefunction
may bifurcate on different surfaces.\cite{Tully1990,Bittner1995,Schwartz1996,Prezhdo1997}
Before the bifurcation event, each FSSH trajectory carries a particular
electronic amplitude. After the trajectory passes through the coupling
region, the wavefunction retains its phase and the density matrix
remains pure, even if the trajectories are separated on different
surfaces. This failure to incorporate decoherence may lead to an inaccurate
description of electronic dynamics. 

The augmented FSSH (A-FSSH)\cite{Landry2012} has been proposed to
resolve this problem by collapsing the electronic state on the inactive
surfaces and projecting onto the active surface according to a decoherence
rate calculated on the fly. The full procedure of the A-FSSH algorithm
is outlined in Ref.~\onlinecite{Landry2012}. Here, for completeness,
we briefly review the A-FSSH scheme.

The decoherence rate depends on the matrix of augmented moments of
the bath coordinate and momentum $\left(\delta\hat{\boldsymbol{q}},\delta\hat{\boldsymbol{p}}\right)$
which provide information regarding the separation of a proxy wave
packet in phase space. The augmented moments evolve along a trajectory
which follows the equations of motion 
\begin{equation}
\frac{d}{dt}\delta\hat{q}_{j}=\hat{T}_{j}^{q}-T_{j,aa}^{q}\hat{I,}
\end{equation}
\begin{equation}
\frac{d}{dt}\delta\hat{p}_{j}=\hat{T}_{j}^{p}-T_{j,aa}^{p}\hat{I},
\end{equation}
where $\hat{T}^{q}$ and $\hat{T}^{p}$ are obtained by expanding
the full quantum Liouville equation to first order in $\hbar$ (linearized
approximation) 
\begin{equation}
\hat{T}_{j}^{q}\equiv-i\left[\hat{V},\delta\hat{q}_{j}\right]+\delta\hat{p}_{j}-\sum_{k}p_{k}\left[\hat{d}_{k},\delta\hat{q}_{j}\right],
\end{equation}
\begin{equation}
\hat{T}_{j}^{p}\equiv-i\left[\hat{V},\delta\hat{p}_{j}\right]+\frac{1}{2}\left\{ \delta\hat{F}_{j},\hat{\sigma}\right\} -\sum_{k}\hat{p}_{k}\left[\hat{d}_{k},\delta\hat{p}_{j}\right],
\end{equation}
and the matrix of forces is given by $\hat{F}_{j}\equiv-\partial\hat{V}/\partial q_{j}|_{\boldsymbol{q}}$
and $\delta\hat{F}_{j}=\hat{F}_{j}-F_{j,aa}\hat{I}$. Via the augmented
moments, one can derive the off-diagonal correction to the equation
of motion for the reduced electronic density matrix, 
\begin{equation}
\frac{d}{dt}\hat{\sigma}=-i[\hat{V},\hat{\sigma}]-\left[\boldsymbol{p\cdot}\hat{\boldsymbol{d}},\hat{\sigma}\right]+i\left[\hat{\boldsymbol{F}},\delta\hat{\boldsymbol{q}}\right],
\end{equation}
which incorporates the decoherence mechanism in the last term. The
estimated decoherence rate for the separation of wavepackets on the
active surface $a$ and the inactive surface $b$ is of the form 
\begin{equation}
\gamma_{ba}^{\text{d}}=dt\left\{ \frac{\left(\boldsymbol{F}_{bb}-\boldsymbol{F}_{aa}\right)\cdot\delta\boldsymbol{q}_{bb}}{2}-2\left|\boldsymbol{F}_{ab}\cdot\delta\boldsymbol{q}_{bb}\right|\right\} ,
\end{equation}
which is obtained by assuming frozen Gaussian wave packets for the
bath wavefunction outside of the derivative coupling region ($\hat{\boldsymbol{d}}=0$)
and reducing the decoherence rate for non-zero derivative couplings.
The A-FSSH algorithm also permits resetting the augmented moments
to avoid the failure of the linearized approximation. The proposed
reset rate is given by bifurcate
\begin{equation}
\gamma_{ba}^{\text{r}}=-dt\frac{\left(\boldsymbol{F}_{bb}-\boldsymbol{F}_{aa}\right)\cdot\delta\boldsymbol{q}_{bb}}{2}.
\end{equation}
Note that $\gamma_{ba}^{\text{r}}$ is the negative collapsing rate
since the moments become invalid when wavepackets aggregate.

A more traditional approach to decoherence corrections within surface
hopping consists of damping the coherence of the density matrix via
a pure-dephasing-like rate.\cite{Prezhdo1998,Zhu2004,Zhu2004a} Within
this simpler density-matrix approach, we treat the evolution of the
adiabatic coherence outside the derivative coupling region ($\hat{\boldsymbol{d}}=0$)
as pure dephasing in a stochastic formulation.\cite{Skinner1986}
Inside the zero derivative coupling region, the population transfer
is excluded and the adiabatic coherences satisfy $\frac{d}{dt}\sigma_{jk}=-i(V_{j}-V_{k})\sigma_{jk}$
and the formal solution is $\sigma_{jk}\left(t+\tau\right)=\sigma_{jk}\left(t\right)\langle\exp\{-i\int_{t}^{t+\tau}dt'[V_{j}\left(t'\right)-V_{k}\left(t'\right)]\}\rangle_{w}$.
The pure-dephasing time within this stochastic formulation is obtained
via the energy difference correlation function\cite{Skinner1986}
\begin{equation}
\frac{1}{T_{2}^{*}}=\frac{1}{2}\int_{0}^{\infty}\left\langle \left[V_{j}\left(t'\right)-V_{k}\left(t'\right)\right]\left[V_{j}\left(0\right)-V_{k}\left(0\right)\right]\right\rangle dt'.\label{eq:dephase_time_1}
\end{equation}
To simulate the decay of the adiabatic coherence within the FSSH algorithm,
we introduce a decoherence terms that leads to an exponential decay
of the adiabatic coherences. In particular, decoherence is modeled
as a Poisson process with the probability that a coherence decay occurs
in the time interval $\left[t,t+dt\right]$ gives by $\text{Prob}\{N[\sigma_{jk}\left(t\right)]-N[\sigma_{jk}\left(t+dt\right)]=1\}=e^{-dt/T_{2}^{*}}dt/T_{2}^{*}\approx dt/T_{2}^{*}$
where $N[\sigma_{jk}\left(t\right)]$ is the number of trajectories
whose density matrix retains coherence. However, for the spin-boson
model, estimation of $T_{2}^{*}$ along each trajectory via Eq.~\eqref{eq:dephase_time_1}
is not well defined. To circumvent this problem, we assume the decoherence
time scale takes a similar form for each trajectory 
\begin{equation}
\frac{1}{\tau_{jk}\left(t\right)}=\frac{1}{2}\int_{0}^{t}\left(V_{j}\left(t'\right)-V_{k}\left(t'\right)\right)\left(V_{j}\left(0\right)-V_{k}\left(0\right)\right)dt'\label{eq:dephase_time_2}
\end{equation}
which gives an estimate of the pure-dephasing time outside of the
derivative coupling region. The decoherence rate for the off-diagonal
term $\sigma_{jk}$ is then given by 
\begin{equation}
\gamma_{jk}^{\text{d}}\left(t\right)=\frac{dt}{\tau_{jk}\left(t\right)}.\label{eq:dephase_rate}
\end{equation}
A decoherence factor for the off-diagonal density matrix elements
may be defined as $\sigma_{jk}=\eta_{jk}c_{j}c_{k}^{*}$, so that
the hopping rate, namely the analogy of Eq.~(\ref{eq:hopping_rate}),
becomes 
\begin{equation}
\gamma_{ab}^{\text{hop}}=dt\frac{2}{|c_{a}|^{2}}\text{Re}(\boldsymbol{p\cdot d}_{ab}\sigma_{ab}).\label{eq:dephase_hopping}
\end{equation}
For every time step, we calculate the decoherence timescale $\tau_{jk}\left(t\right)$
by accumulating energy difference correlations along the trajectory.
If a decoherence event occurs, the associated factor $\eta_{jk}$
is set to zero. Then we symmetrize the density matrix and continue
the trajectory propagation.

\begin{figure}
\includegraphics{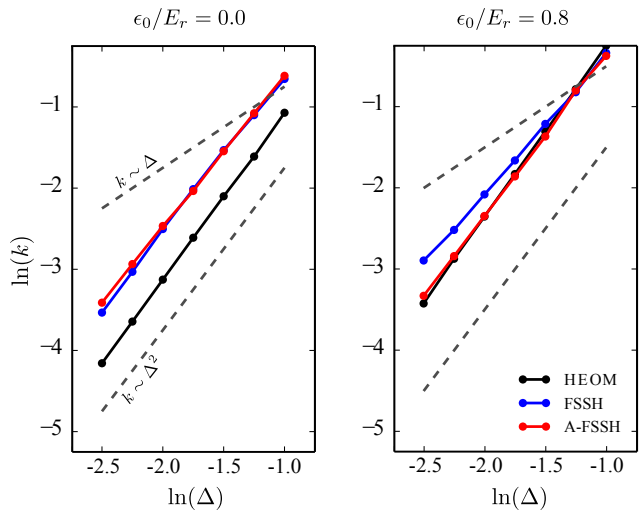}

\caption{Diabatic population transfer rates ($k$) as a function of diabatic
coupling $\Delta$ for FSSH (blue), A-FSSH (red), and HEOM (black)
in the unbiased $\epsilon_{0}/E_{r}=0$ and biased $\epsilon_{0}/E_{r}=0.8$
cases. The bath temperature is assumed to be in the classical limit,
$T=300\ \text{K}$. The reorganization energy is $E_{r}=520\ \text{cm}^{-1}$,
while the bath frequency scale $\omega_{c}$ is tuned so that $\Delta/\omega_{c}\ll1$.
The dashed lines are reference markers of sub-quadratic and quadratic
dependence, respectively. The diabatic population transfer rates is
extracted from the population dynamics by exponential fitting. \label{fig:lnk-lnD}}
\end{figure}

\begin{figure}
\includegraphics[clip]{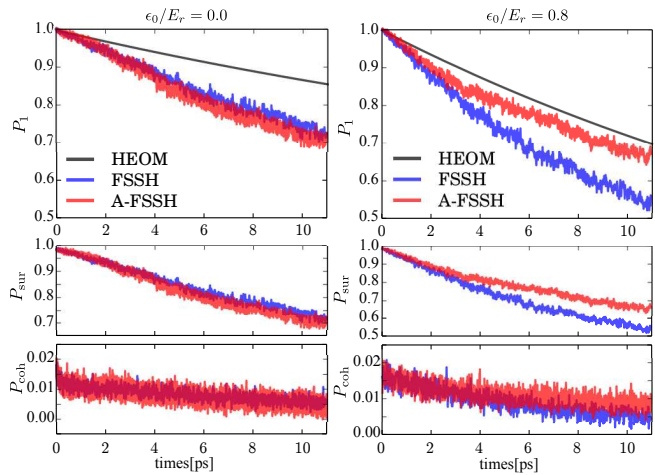}

\caption{Population dynamics of the for FSSH, A-FSSH, and HEOM in the (a) unbiased
$\epsilon_{0}/E_{r}=0$ and (b) biased $\epsilon_{0}/E_{r}=0.8$ cases.
The lower panels show the surface and coherence terms separately,
as defined in Eq.~(\ref{eq: surface_and_coherence}). The bath temperature
is $T=300\ \text{K}$, the reorganization energy is $E_{r}=520\ \text{cm}^{-1}$,
and $\omega_{c}=85\ \text{cm}^{-1}$.\label{fig:pop-Marcus}}
\end{figure}

\section{The Golden-Rule Regime\label{sec:The-Golden-Rule}}

A surprising feature of the standard FSSH algorithm that has recently
been discovered is its failure to capture the quadratic dependence
of the (diabatic) transfer rate in the weak coupling regime. The generality
of this behavior remains somewhat controversial. Furthermore, the
fundamental origin of this apparent failure is unclear. Landry and
Subotnik noted an interesting aspect of the simple one-dimensional
Landau-Zener (LZ) problem.\cite{Landry2011} In the standard treatment
of the LZ problem with initial electronic population on one surface
only, a single voyage through the crossing region produces population
differences in harmony with the expected quadratic coupling dependence
of the rate. However, if the system is prepared initially with arbitrary
population on \textit{both} diabatic surfaces, then a passage through
the crossing point induces a population change that is proportional
to \textit{both} the electronic coupling itself as well as its square.
It may then be argued that since traversal of the crossing region
mixes the populations, multiple crossings will produce a rate with
a sub-quadratic coupling dependence. With the addition of decoherence,
however, populations are localized after each crossing, such that
the rate always retains its proper quadratic golden-rule form. Below
we show that while this argument \textit{cannot} explain the deviations
from Marcus golden-rule behavior exhibited by FSSH, the notion that
decoherence can alter the coupling dependence of the rate in a favorable
way is indeed correct.

Let us briefly revisit the simple one dimensional LZ example. As in
Ref.~\onlinecite{Landry2011}., let us take an electronic propagator
of the form $U=\left(\begin{array}{cc}
\sqrt{1-\xi} & \sqrt{\xi}\\
-\sqrt{\xi} & \sqrt{1-\xi}
\end{array}\right)$, where $\xi=\exp\left[-\frac{2\pi}{\hbar}\frac{\Delta^{2}}{\left|\boldsymbol{v}\cdot\left(\boldsymbol{F_{1}-F_{2}}\right)\right|}\right]\equiv\exp\left[-\eta\Delta^{2}\right]$
is the LZ parameter which depends on the crossing velocity $\boldsymbol{v}$
and the difference in the (diabatic) forces, $\boldsymbol{F_{1}-F_{2}}$,
at the crossing point, and the electronic coupling, $\Delta$. Clearly
a pure initial wave packet with amplitude placed entirely on surface
$a$, namely $P\left(0\right)=\left(\begin{array}{c}
1\\
0
\end{array}\right)$, produces a population difference on surface $b$ after one crossing
that is proportional to $\Delta^{2}$ for small $\Delta$. On the
other hand, if the initial packet has the form $P\left(0\right)=\left(\begin{array}{c}
\alpha\\
\beta
\end{array}\right)$ where $\alpha$ and $\beta$ are arbitrary constants satisfying $\alpha^{2}+\beta^{2}=1$,
then after one passage the population difference on the surface $b$
is given by $(\xi-1)\beta^{2}+(1-\xi)\alpha^{2}-2\sqrt{\xi(1-\xi)}\alpha\beta\approx\left(\alpha^{2}-\beta^{2}\right)\eta\Delta^{2}-\sqrt{\eta}\Delta\alpha\beta$.
The linear term in the electronic coupling heralds an apparent subquadratic
dependence of the rate on $\Delta$. Importantly, however, it should
be noted that the mixing of populations that occurs during passage
through the crossing region depends on $\Delta$. In particular, starting
from the \textquotedbl{}pure\textquotedbl{} initial state $P\left(0\right)=\left(\begin{array}{c}
1\\
0
\end{array}\right)$, passage through the crossing region produces populations on each
diabatic state that are non-zero, but do depend on $\Delta$ and are
thus \textit{not} arbitrary constants. Via consideration of $U^{n}\left(\begin{array}{c}
1\\
0
\end{array}\right)$, it is straightforward to demonstrate that even in the absence of
decoherence, multiple crossings do not generate spurious terms in
the $b$-state population that are linear in $\Delta$ within this
simple model.

To explore the issue of the behavior predicted by surface hopping
in the Marcus regime, we turn to direct simulation. In Fig.~\ref{fig:lnk-lnD},
compare the exact diabatic population transfer rates, numerically
extracted from HEOM simulations in the high temperature, weak electronic
coupling regime to both the results predicted by FSSH as well as the
decoherence based A-FSSH algorithm. In both cases, we use Eq.~(\ref{eq:P1_expression})
to extract diabatic quantities. The exact HEOM simulations are not
confined to the strict high temperature limit. Thus we expect rates
that scale as $\Delta^{2}$, but do not necessarily conform quantitatively
to standard Marcus theory. The results are shown for both an unbiased
and strongly biased cases of the spin-boson problem. Several important
features should be noted. First, in the symmetric situation, the FSSH
approach yields the correct scaling of the rate with $\Delta$ and
produces results that are essentially indistinguishable from those
of A-FSSH. This is true even as the electronic coupling is varied
over a wider range, and for all values of the reorganization energy.
On the other hand, when there is a sizable energetic bias, the rate
indeed violates Marcus scaling and behaves in a manner qualitatively
similar to that described in Ref.~\onlinecite{Landry2012}.\footnote{The recently published paper, Jain and Subotnik, J. Phys. Chem. Lett.
\textbf{6} , 4809 (2015), makes a nearly identical observation. We
thank Joseph Subotnik for making us aware of this during the writing
of this manuscript.} \textit{Importantly, however, the magnitude of the deviations we
find are significantly smaller than that expected from the calculations
of Ref.~\onlinecite{Landry2012}.} Remarkably, the inclusion of
decoherence corrects this failing, producing results in quantitative
correspondence with exact numerics. Thus, violations of the expected
golden-rule behavior as well as the impact of decoherence in the weak-coupling
regime appear to depend sensitively on the electronic bias.

To gain a deeper understanding of this surprising result, we decompose
the non-adiabatic population into terms that have an explicit dependence
on the dynamics on a given surface the the coherence between surfaces,
respectively. It may be shown that Eq.~(\ref{eq:P1_expression})
can be recast as 
\begin{eqnarray}
P_{1} & = & P_{\text{sur}}+P_{\text{coh}}\nonumber \\
 & = & \left\langle \frac{1}{2}+\frac{1}{2}\frac{\boldsymbol{g}\cdot\boldsymbol{q}+\epsilon_{0}}{\sqrt{(\boldsymbol{g}\cdot\boldsymbol{q}+\epsilon_{0})^{2}+\Delta^{2}}}(\delta_{2\lambda}-\delta_{1\lambda})\right\rangle \nonumber \\
 &  & +\left\langle \frac{\Delta}{\sqrt{(\boldsymbol{g}\cdot\boldsymbol{q}+\epsilon_{0})^{2}+\Delta^{2}}}\text{Re}[c_{1}c_{2}^{*}]\right\rangle ,\label{eq: surface_and_coherence}
\end{eqnarray}
where we have labeled the two relevant terms in Eq.~(\ref{eq: surface_and_coherence})
as the \textquotedbl{}surface\textquotedbl{} term, $P_{\text{sur}}$,
and the \textquotedbl{}coherence\textquotedbl{} term, $P_{\text{coh}}$.
Note that we are using the diabatic interpretation of Ref.~\onlinecite{Landry2013},
so in essence it is the \textquotedbl{}surface\textquotedbl{} term
that is expected to be most sensitive to decoherence corrections applied
in the \textit{adiabatic} basis, not the \textquotedbl{}coherence\textquotedbl{}
term. Furthermore, note that it is the surface term that has the stronger
explicit dependence on the energetic bias, in harmony with the notion
that the distinction between FSSH and its decoherence corrected variants
will depend on bias as reflected in the way decoherence alters the
behavior of the first term of Eq.~(\ref{eq:P1_expression}). In Fig.~\ref{fig:pop-Marcus}
we show the temporal decay of population in both the unbiased and
biased cases, within both FSSH and A-FSSH. We also show separately
the surface and coherence terms. For the unbiased case, FSSH and A-FSSH
yield essentially identical results, while in the biased case A-FSSH
is in near quantitative agreement with the exact result while the
standard FSSH result decays too rapidly. The difference between the
two results is noticeable only in the surface term, which dominates
over the coherence term. Thus, we find that distinction between the
unbiased and biased cases reflects the manner in which the bias couples
to coherence-sensitive terms as exposed in Eq.~(\ref{eq: surface_and_coherence}).

\section{the full parameter space\label{sec:the-full-parameter}}

\begin{figure}
\begin{raggedright}
(a) 
\par\end{raggedright}

\includegraphics[scale=0.95]{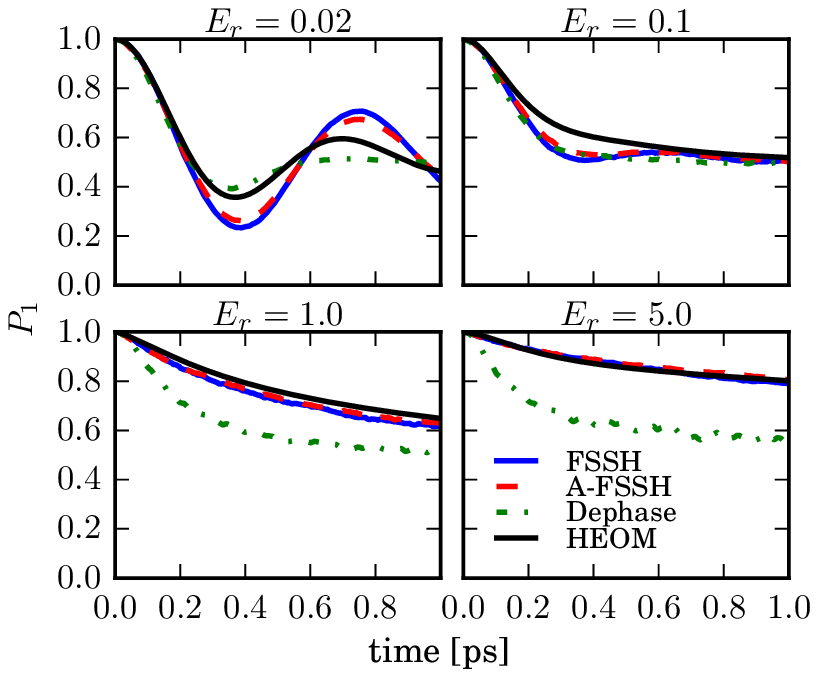}

\vspace{-10bp}

\begin{raggedright}
(b) 
\par\end{raggedright}

\includegraphics[scale=0.95]{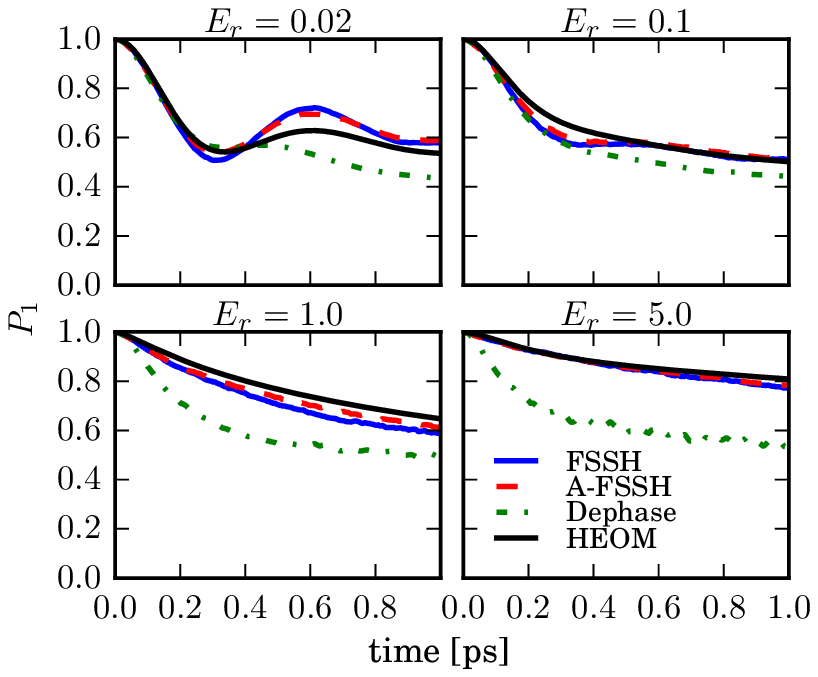}

\vspace{-10bp}

\begin{raggedright}
(c) 
\par\end{raggedright}

\includegraphics[scale=0.95]{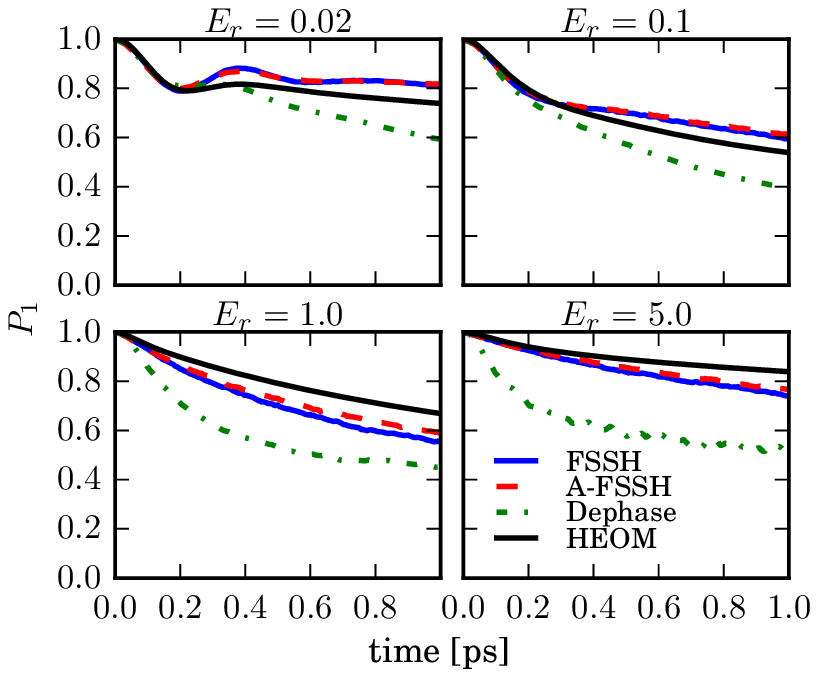}

\caption{High temperature dynamics with intermediate electronic coupling strength.
We employ a reference unit of energy of $104\ \text{cm}^{-1}$. Parameters
are $kT=2$ ($T=300\ \text{K}$), $\Delta=\omega_{c}=0.2$ , (a)$\epsilon_{0}/\Delta=0$,
(b)$\epsilon_{0}/\Delta=2$, and (c)$\epsilon_{0}/\Delta=4$. Reorganization
energies are scanned from small to large and are listed on each panel.
``Dephase'' refers to the use of Eqs.~\eqref{eq:dephase_time_2}-\eqref{eq:dephase_hopping}.}
\end{figure}

\begin{figure}
\begin{raggedright}
(a) 
\par\end{raggedright}

\includegraphics[scale=0.95]{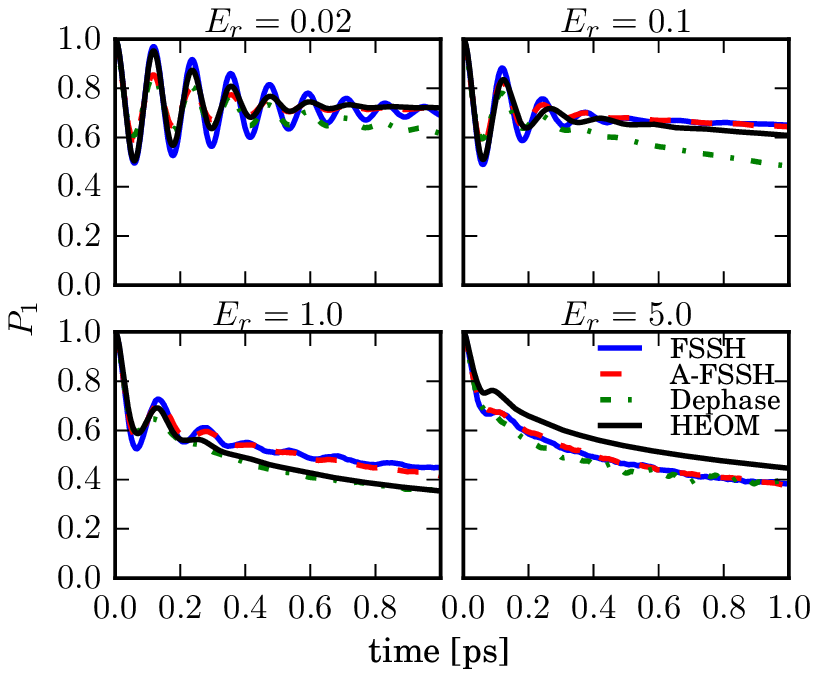}

\vspace{-10bp}

\begin{raggedright}
(b) 
\par\end{raggedright}

\includegraphics[scale=0.95]{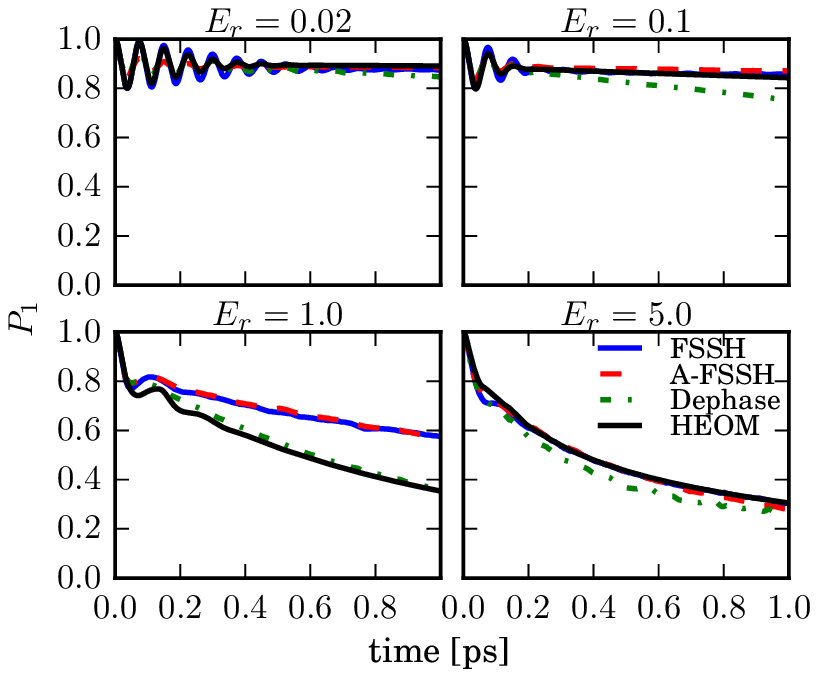}

\caption{High temperature dynamics in the adiabatic regime. We employ a reference
unit of energy of $104\ \text{cm}^{-1}$. Parameters are $kT=2$ ($T=300\ \text{K}$),
$\Delta=1$, $\omega_{c}=0.2$, (a)$\epsilon_{0}/\Delta=2$ and (b)$\epsilon_{0}/\Delta=4$.
Reorganization energies are scanned from small to large and are listed
on each panel. ``Dephase'' refers to the use of Eqs.~\eqref{eq:dephase_time_2}-\eqref{eq:dephase_hopping}.}
\end{figure}

\begin{figure}
\includegraphics[scale=0.95]{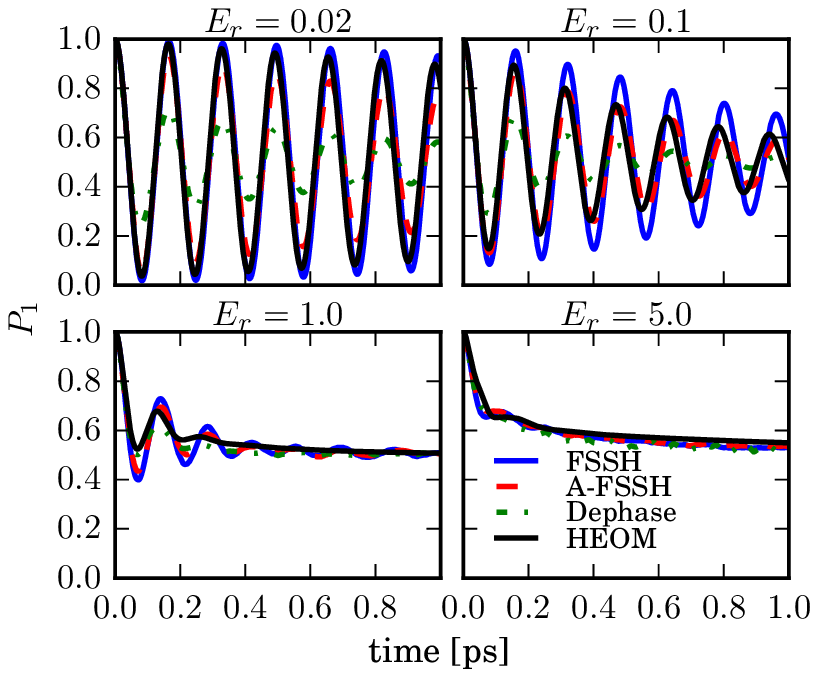}

\caption{High temperature dynamics in the adiabatic regime. We employ a reference
unit of energy of $104\ \text{cm}^{-1}$. Parameters are $kT=2$ ($T=300\ \text{K}$),
$\Delta=1$, $\omega_{c}=0.2$, and $\epsilon_{0}=0$. Reorganization
energies are scanned from small to large and are listed on each panel.
``Dephase'' refers to the use of Eqs.~\eqref{eq:dephase_time_2}-\eqref{eq:dephase_hopping}.}
\end{figure}

In this section, we explore more broadly the comparison of surface
hopping to benchmark calculations of dynamics in the spin-boson model.
Fig.~\ref{fig:pop-Marcus} illustrates that in the golden-rule regime,
standard FSSH produces results in qualitative agreement with the exact
behavior produced by HEOM calculations. The inclusion of decoherence
can lead to improved and even quantitatively accurate results, however
the improvement over FSSH will depends sensitively on the parameters
of the underlying Hamiltonian, such as the energetic bias. Similar
behavior is seen away from the weak coupling limit. For intermediate
electronic coupling and high temperature ($kT/\omega_{c}\gg1$), a
regime often difficult to treat via approximate perturbative approaches,
we find that FSSH is quite accurate, with an accuracy that is not
altered by inclusion of decoherence within the A-FSSH approach. On
the other hand, direct decoherence damping with a pure-dephasing-type
rate generally leads to less accurate results than FSSH in this regime,
especially when the reorganization energy is large. These observations
are illustrated in Fig.~3. In the adiabatic regime, where the electronic
coupling is large, we again find that FSSH is in good agreement with
the exact behavior of the simulated non-equilibrium populations at
high temperatures, especially for large reorganization energies. Some
select examples of this comparison are illustrated in Fig.~4. In
situations where the reorganization energy is small and the system
has no energetic bias, the upper left panel of Fig.~4(a) and the
upper right panel of Fig.~5 illustrate how A-FSSH provides a damping
of population oscillations that brings the approximate results into
quantitative correspondence with exact simulations. With respect to
more phenomenological treatments of decoherence, two new features
stand out. First, direct decoherence damping with a pure-dephasing
rate generally leads to more accurate results in the strong-coupling
regime than it does in situations where the electronic coupling is
intermediate or small as compared to other energy scales in the problem.
In particular, unlike in the case of intermediate coupling, the more
phenomenological treatment of decoherence appears not to lead to gross
overestimates of the rate of population decay in energetically biased
cases for large electronic couplings. Furthermore, we find, for the
first time, examples where a simple \textquotedbl{}pure dephasing\textquotedbl{}
correction leads to clearly improved accuracy over both FSSH and A-FSSH.
We emphasize however that in general we find A-FSSH to be, on average,
the most accurate approach across the full parameter space.

Lastly, we turn to situations where the temperature is comparable
to, or lower than, the characteristic bath frequency. In such situations
we expect any surface hopping approach to be unreliable due to the
fact that the dynamics of the nuclei are treated classically. Thus
processes such as nuclear quantum tunneling cannot be described. While
we find this to be generally the case, there are situations where
the surface hopping approaches find some success even in this \textquotedbl{}quantum
bath\textquotedbl{} regime. In particular, when the electronic coupling
is strong and the time scale is relatively short, both FSSH and A-FSSH
can accurately model the Rabi-like oscillations for several periods
of motion as illustrated in the upper left panel of Fig.~6 (a) and
(b). For intermediate temperatures A-FSSH can accurately correct the
decay rate of the amplitude of oscillations, however its accuracy
diminishes at lower temperatures as shown in Fig.~7. In general,
however, surface hopping fails to quantitatively capture population
relaxation in these regimes, with some \textquotedbl{}worst-case\textquotedbl{}
examples illustrated in Fig.~6.

\begin{figure}
\begin{raggedright}
(a)
\par\end{raggedright}

\includegraphics[scale=0.95]{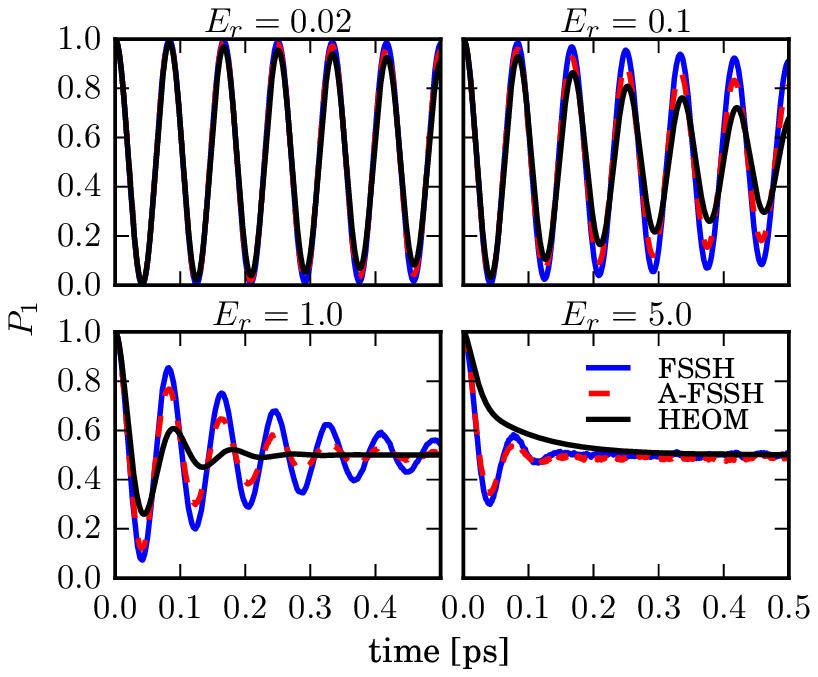}

\vspace{-10bp}

\begin{raggedright}
(b) 
\par\end{raggedright}

\includegraphics[scale=0.95]{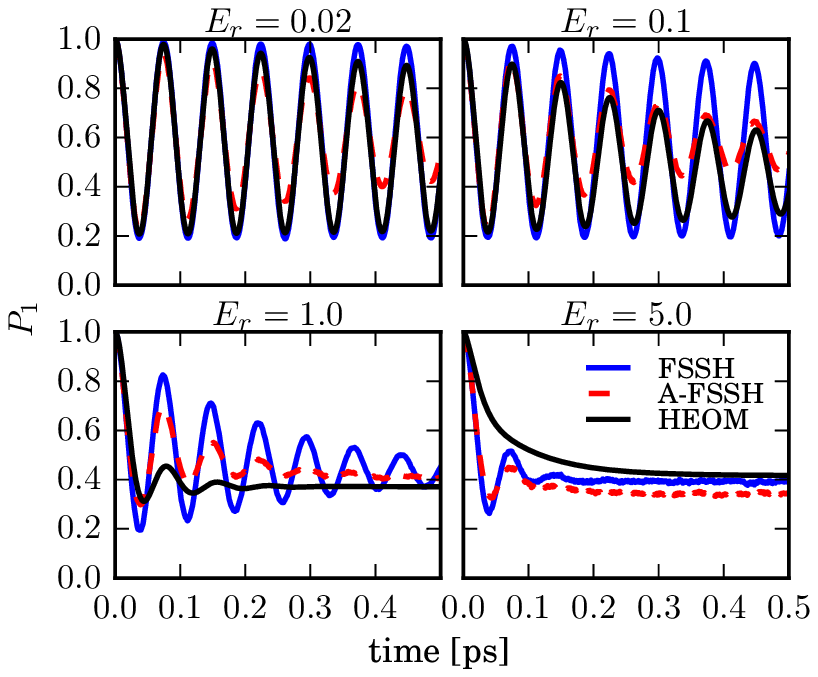}

\caption{Low temperature dynamics in the intermediate regime. We employ a reference
unit of energy of $104\ \text{cm}^{-1}$. Parameters are $kT=0.2$
($T=30\ \text{K}$), $\Delta=2$, $\omega_{c}=2$, $kT=0.2$, (a)
$\epsilon_{0}/\Delta=0$, and (b)$\epsilon_{0}/\Delta=2$. Reorganization
energies are scanned from small to large and are listed on each panel.}
\end{figure}

\begin{figure}
\includegraphics[scale=0.95]{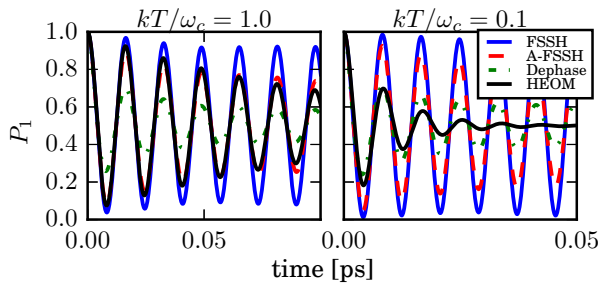}

\caption{Intermediate and low temperature dynamics in the adiabatic regime.
We employ a reference unit of energy of $104\ \text{cm}^{-1}$. $\epsilon_{0}=0$
and reorganization energy is large $E_{r}=5$. Parameters are (left)
$\Delta=10$, $\omega_{c}=1$, $kT=1$ and (right) $\Delta=20$, $\omega_{c}=2$,
$kT=0.2$. ``Dephase'' refers to the use of Eqs.~\eqref{eq:dephase_time_2}-\eqref{eq:dephase_hopping}.}
\end{figure}

\section{Conclusions\label{sec:Conclusions}}

In this work we have provided, to the best of our knowledge, the first
detailed comparison of surface hopping with exact quantum dynamics
for an idealized but non-trivial model of condensed phase non-adiabatic
dynamics. In particular, we have focused on the role played by decoherence
across the entire parameter space in general, and in the incoherent
golden-rule regime in particular. Our results provide both an understanding
of how decoherence influences behavior in the weak electronic coupling
regime as well as general guidelines for the reliability of surface
hopping with or without decoherence corrections across all regimes.

With respect to recovery of Marcus golden-rule scaling behavior, we
present several novel findings. First, we find that deviations from
golden-rule scaling, at least within the confines of the spin-boson
model with a standard Debye spectral density, do not occur for symmetric
systems and only become apparent in systems with a large energetic
bias. In biased cases the inclusion of decoherence appears to correct
the errant behavior of the standard FSSH approach. On the other hand,
we show that the origins of the inability of FSSH to yield golden-rule
behavior are subtle and the departure from the quadratic scaling are
smaller than expected from past work. Lastly, we note that while it
is clear that the decoherence based A-FSSH algorithm alters the electronic
dependence of the transfer rate in the weak coupling limit so that
the standard golden-rule is recovered, we have no analytical argument
that demonstrates that this should occur, or that it will continue
to be true over a wider range of $\Delta$ than we have investigated. 

A systematic survey of parameter space provides important guidelines
concerning the accuracy of surface hopping and its decoherence-corrected
variants. One major conclusion that can be immediately reached is
that, in general, the standard FSSH is surprisingly accurate in large
portions of parameter space. Furthermore, while the decoherence-based
A-FSSH approach often leads to some improvement in the description
of the temporal decay of non-equilibrium population, on average the
corrections are not dramatic. The largest improvements fostered by
the inclusion of decoherence provided within the A-FSSH approach are
found in the previously discussed golden-rule regime (c.f. Fig.~2)
as well as in cases where decoherence damps otherwise oscillatory
population decay. Thus, at least with respect condensed phase environments
with widely dispersed spectral properties, the standard FSSH approach
should generally provide a reasonable description of dynamics.

All of the surface hopping approached we have employed in this work
have difficulty in accurately describing low temperature situations,
with the exception of symmetric cases where the electronic coupling
is so weak that essentially pure Rabi oscillations are observed on
short to intermediate time scales. However this breakdown of surface
hopping is unsurprising as the approach is incapable of capturing
nuclear tunneling effects. Quantitative breakdowns also appear at
high and intermediate temperatures not only in the golden-rule limit,
but also for intermediate to strong electronic coupling when the coupling
to the bath (as given in the reorganization energy) is also sizable.
However, even in these regimes failures appear as isolated examples
more than generic trends.

We have also investigated decoherence corrections that are perhaps
less well justified than that provided by A-FSSH but are simpler conceptually.
In particular, we have explored an approach similar to the earliest
decoherence corrections which employs a simple damping term given
by the pure dephasing rate along a trajectory. In general we find
that such an approach decoheres relaxation dynamics too strongly,
often worsening agreement between the standard FSSH algorithm and
the exact results. Somewhat surprisingly however, the degree of decoherence
provided by this approach may be seen to quantitatively correct the
failures of both FSSH and A-FSSH in the \textquotedbl{}isolated\textquotedbl{}
cases where both fail, namely the regimes of sizable electronic and
system-bath couplings mentioned above. This coincidence should be
investigated further, as it may foster a deeper understanding of the
physics associated with these isolated examples, something that we
currently have been unable to provide.
\begin{acknowledgments}
We would like to thank Joseph Subotnik, Brian Landry, Thomas Markland,
Aaron Kelly, and Andrés Montoya-Castillo for extensive discussions.
This work was supported by grant NSF CHE-1464802.
\end{acknowledgments}

\bibliographystyle{apsrev4-1}
\bibliography{FSSH}

\end{document}